\documentstyle[twocolumn,prc,aps,epsf,epsfig]{revtex}
 \begin{document}
 
\newcommand{\be}{\begin{eqnarray}}
\newcommand{\ee}{\end{eqnarray}}
\twocolumn[\hsize\textwidth\columnwidth\hsize\csname@twocolumnfalse\endcsname
\title{A Cosmological Constant  from Gauge Field Instantons? 
}
\author{ E.V. Shuryak }
\address{
Department of Physics and Astronomy, State University of New York, 
Stony Brook NY 11794-3800, USA
}

\date{\today}
\maketitle

\begin{abstract}
Although all interactions in the Standard Model generate nonzero shifts
of the vacuum energy and pressure, 
 gravity does not interact with them. Assuming (i) that the reason why
 it is so
breaks down at some scale $M_g$ and that (ii) the instanton-induced shifts
at such scale
generate the observed cosmological constant, we found that it
then should happen at a (surprisingly
small) scale   $M_g\sim 10^3 TeV$.
\end{abstract}
\vspace{0.1in}
]
\begin{narrowtext}
\newpage
  Recent observations of dark energy and accelerated expansion of
Universe, combined with deceleration in the  past  have
provided strong indication for possible existence of the cosmological
constant, for overview see \cite{Weinberg:2000yb}. Its magnitude at present is about .7 of the critical energy
density.

As the
 nature of this ``dark energy''
remains mysterious, even quite speculative possible relations between
its scale 
\be M_\Lambda=\Lambda^{1/4} \approx 110 cm^{-1} \approx  0.002 eV \ee
 (where $\Lambda$ is the dark energy density)
to fundamental parameters had generating interesting debates.

A simple example of such relations is 
 a {\em geometric mean}
between the Plank mass and the inverse current Universe size, see
\cite{U_size}).
However, in spite of its elegant simplicity, 
 it is hard to imagine any dynamical mechanism involving an infrared cutoff
of the order of the size of Universe. In this note we will hold more
traditional view, that the resolution should resign somewhere at
 higher
 energies.

Another notable fact is that   $M_\Lambda$ happens to be
 comparable to the scale
of the neutrino masses, and (if not accidental) it hints that
 some
dynamics ``beyond the Standard Model'' may be involved.

  As a motivation to the relation we propose, we would like to stress
first that the power-like combinations of parameters usually used
in such relations are not inevitable, and we know of many examples
when exponential (or logarithmic) behavior leads to unusually
large/small
new scales. The closest to our proposal is the unusually long lifetime
of alpha-radioactive nuclei, which can be enhanced
be nearly 40 orders of magnitude
 in time scales  by  exponentially small
tunneling suppression, the semiclassical Gamow factor
\be P_{Gamow}\sim exp \left[-{2\pi (2e^2) (Z-2)\over hv}\right]\ee
with the velocity $v$ as the small parameter,  $2(Z-2)$ are  the
 charges of the alpha-particle and the remaining
 nucleus\footnote{Plank constant $h$ in this formula will not appear
in other expressions, where we use the usual $\hbar=c=1$ units.}.

  Instantons \cite{BPS_75} are also tunneling events, with a
  similar
exponential factor given by the action $exp(-8\pi^2/g^2)$ with the
  gauge
coupling being a small parameter. Furthermore, due to topology involved,
chirality of fermions get into the game and
  the fermionic determinant appear,  producing another very small
  factor \cite{tHooft}.
Their multiple uses in strong
  interactions are well known (for a review see \cite{SS_98}). 

 Instantons  produce a nonperturbative shift of the vacuum energy and pressure.
The energy density due to strong instantons is at a scale
of $B_{strong}\sim 1 \, GeV/fm^3$. As instantons get
suppressed in  quark-gluon plasma, this shift is acting like a
``dark energy'' in the ``Little Bangs'' of heavy ion collisions,
decelerating expansion (see more on that in \cite{hydro}).
Weak instantons were much discussed as a source of a baryon number
  violation in the 1990's, 
but they seem to produce too tiny effect to be relevant,  even for
  cosmological constant (see below).

  For some unknown reasons, this vacuum energy  $B_{strong}$ and
even much large one $B_{weak}$ (related to electroweak
Higgs phenomenon) does not
interact with gravity, so to say it has {\em zero weight}. All vacuum
contributions of all orders as well as the exponential effects
(instantons) get canceled out and somehow become
 invisible to gravity. 

  One may still expect, that this
mechanism gets broken  at some high scale $M_g$
where  the laws of gravity are changed. The usual Einstein
gravity suggests it to happen at the Plank mass scale, $10^{19} GeV$,
 but other (nowadays quite
popular) ideas about extra dimensions had shown that it may happen
at scale as low as a $TeV$ \cite{Arkani-Hamed:1998nn}.

The main assumption made in this short note
is that where the gravity laws change, {\em they only allow the  exponential
instanton terms} not to cancel and thus
provide a nonzero weight to the vacuum\footnote{ The
 power terms should still be canceled, preventing huge
cosmological
constant to appear, which we know is phenomenologically absent 
in our Universe. Some supersymmetric models show examples of that.}.
The main question we address then is {\em at what scale should it happen},
so that the observed cosmological constant scale $M_\Lambda$ be reproduced.

 It is instructive to answer this question in two steps, starting first with
a rather naive   estimate with only exponential term included 
\be \Lambda \sim M^4 exp\left(-{8\pi^2\over g^2(M)}\right)\ee
If so, one gets an impression that $M_g$ in question may be a 
 very high scale like the Plank mass or Grand 
Unification scale $M_U=10^{16} GeV$. Indeed, taking the latter
as an example  one finds that if
the exponent provides suppression by 110 orders of magnitude,
the corresponding coupling is  $(g^2/4\pi)^{-1} \approx 40$ which is
indeed close to a value  expected for the unified gauge 
coupling at this scale\footnote{  E.g.
the  original 
unification paper by Georgi,Quinn and Weinberg \cite{GQW} mentioned (in crude
  one-loop
approximation) at unification scale
$(g^2/4\pi)^{-1}\sim 48$ and huge further literature on improved beta function
 and 
SUSY extensions.}.

The second step includes the additional  suppression
due to non-exponential factors originating from
 bosonic and fermionic zero modes. 
The fermions (and their masses) involved depend on a specific orientation
of the SU(2) subgroup, in which
instanton is located, inside the unified theory. However, since in each
 generation
quark masses are larger than those of leptons, the instantons of
strong
forces always win
over weak ones, even at the same (unified) gauge coupling. So we rewrite the
 expression above 
as
\be \Lambda \sim  \left[{8\pi^2\over g^2(M)}\right]^6 \Pi_f\left({m_f\over M}\right) M^4 exp(-{8\pi^2\over g^2(M)})\ee
where the first term stands for 12 bosonic zero modes of SU(3) color
group, while  the second 
includes the product of masses
of all 6 ($u,d,s,c,b,t$) quark flavors\footnote{Due to chiral
 symmetry breaking in QCD vacuum, for light $u,d,s$
quarks here enters effective masses $\sim .1 GeV$ proportional to
respective quark condensates.  }. This provide huge small parameter by itself.

Demanding this expression to reproduce the observed cosmological constant
we find that it would happen at the ``gravity breaking'' scale 
\be M_g\approx
10^3 \, TeV .\ee Although still not a $TeV$ scale (soon to be addressed
at LHC),
 it is only 3 orders of magnitude away from current frontier and
much 
closer to observability than the unification or the Plank scale. 
It is not unreasonable to think that one day anomalous 
coupling of gluons to gravitons at such a scale may become testable.

In summary, we suggest that the cosmological constant  may originate from
the vacuum energy shift induced by the usual (gluon gauge field)
instantons beyond a scale where gravity laws change 
and make this shift visible to gravitons, at the scale $M_g\approx
10^6 GeV$. 
 (Needless to say,
we  cannot explain neither the fine tuning mechanism at lower
scale nor what exactly happens there: we just indicate a scale
where it $may$ happen.)

\end{narrowtext}
\end{document}